# Structure-Property Relationship in Layered $BaMn_2Sb_2$ and $Ba_2Mn_3Sb_2O_2$

Qiang Zhang[1,2], Zhenyu Diao[1], Huibo Cao[2], Ahmad Saleheen[1], Ramakanta Chapai[1], Dongliang Gong[1], Shane Stadler[1], R. Jin[1*]

[1]Department of Physics and Astronomy, Louisiana State University, Baton Rouge, LA 70803 USA

[2]Neutron Scattering Division, Oak Ridge National Laboratory, Oak Ridge, TN 37831, USA

**Abstract**

Layered transition-metal compounds have received great attention owing to their novel physical properties. Here, we present the structural, electronic, thermal, and magnetic properties of $BaMn_2Sb_2$ and $Ba_2Mn_3Sb_2O_2$ single crystals, both with the layered structure analogous to high-temperature superconductors. While the Mn moment in the $MnSb_4$ tetrahedral environment forms *G*-type antiferromagnetic (AFM) ordering in both $BaMn_2Sb_2$ ($T_{N1} \approx 443$ K) and $Ba_2Mn_3Sb_2O_2$ ($T_{N1} \approx 314$ K), a short-range AFM order is found in the intercalated $MnO_2$ layer at a much lower temperature ($T_{N2} \approx 60$ K) in $Ba_2Mn_3Sb_2O_2$. The directions of the ordered moments in these two magnetic sub-lattices of $Ba_2Mn_3Sb_2O_2$ are perpendicular to each other, even though the system is electrically conductive. This indicates that the large magnetic moments in these compounds are highly localized, leading to negligible coupling between $MnSb_4$ and $MnO_2$ layers in $Ba_2Mn_3Sb_2O_2$. These findings provide an insight into the structure-magnetism-based design principle for new superconductors.

*Email: rjin@lsu.edu

## I. INTRODUCTION

The discovery of high-transition-temperature (high-$T_c$) superconductivity in $CuO_2$-based[1] and $FeAs_4$-based[2] layered materials has triggered extensive studies of layered transition-metal compounds involving similar crystal structures and building blocks, aimed at searching for possible new superconductors and underlying physics. Compounds with the general formula $AMn_2Pn_2$ (A = an alkaline earth metal; Pn = pnictogen)[3-6] are isostructural to $AFe_2As_2$[2,7], a parent compound of the 122-type Fe-based superconductors. The fundamental building block is the $MnPn_4$ tetrahedron, which forms a layer by edge sharing as shown in Fig. 1(a). In contrast to the metallic behavior in the $AFe_2As_2$ family, Mn-based $BaMn_2Pn_2$ has the semiconducting or insulating ground state[4-6]. Furthermore, $BaMn_2As_2$ was reported to exhibit a *G*-type antiferromagnetic (AFM) order[4], distinct from the stripe-like AFM order in the $AFe_2As_2$ series[2,7]. In addition, the replacement of As by heavier Sb may influence the physical properties of a system. For example, the AFM ordering temperature of CeMnSbO is much lower than that of CeMnAsO[10]. On the other hand, LiFeSb is predicted to have higher $T_c$ than LiFeAs[8,9].

Besides chemical element replacement, another way of designing new superconductors involves the inclusion of both the $FeAs_4$- and $CuO_2$- types building blocks. Compounds with the general formula $A_2Mn_3Pn_2O_2$ are built upon $Mn(1)Pn_4$ and $Mn(2)O_2$ building blocks (see Fig. 1(b)) [11-15], the latter being similar to the $CuO_2$ layer in high- $T_c$ cuprates. $A_2Mn_3Pn_2O_2$ can be viewed as the insertion of a $Mn(2)O_2$ layer in $AMn_2Pn_2$ separated by A. The structural similarity between $AMn_2Pn_2$ and $A_2Mn_3Pn_2O_2$ offers an excellent opportunity to study the structure-property relationship, especially the roles of the $CuO_2$- and $FeAs_4$-type layers in their physical properties. In this article, we focus on the experimental investigation of the structural and physical properties of $BaMn_2Sb_2$ and $Ba_2Mn_3Sb_2O_2$ single crystals. The comparative studies allow us to identify the roles of the Mn(1)Sb$_4$ and Mn(2)O$_2$ layers in these compounds, which will help guide the design of new superconductors.

## II. EXPERIMENTAL DETAILS

Single crystals of BaMn$_2$Sb$_2$ and Ba$_2$Mn$_3$Sb$_2$O$_2$ were grown by the flux method using Sn as flux. For the growth of BaMn$_2$Sb$_2$, Ba rod (Alfa Aesar 99+%), Mn powder (Alfa Aesar, 99.95%), Sb powder (Alfa Aesar, 99.999%), and Sn powder (Alfa Aesar, 99.995%) were mixed in the ratio of 1 : 2 : 2 : 5, and placed in an alumina crucible. The crucible was then sealed in an evacuated quartz tube. The whole assembly was first sintered at 1180 ºC for 15 h, and then slowly cooled down to 700 ºC at a rate of 5 ºC/h. The sealed quartz tube was taken out of the furnace, spun in a centrifuge, and finally quenched to room temperature. The same temperature profile and procedure were used to grow Ba$_2$Mn$_3$Sb$_2$O$_2$ single crystals out of the mixture of BaO : Mn : Sb : Sn with the ratio of 2 : 3 : 2 : 5. In both growths, plate-like single crystals with surfaces over 5 mm × 5 mm were obtained, which are very malleable and easily bent.

The compositions of grown crystals were measured using the wavelength dispersive spectroscopy (WDS) technique. For targeted BaMn$_2$Sb$_2$, the average elemental ratio was found to be Ba : Mn : Sb = 1 : 1.94 : 2, which is consistent with 1 : 2 : 2 stoichiometry with slight Mn deficiency. For targeted Ba$_2$Mn$_3$Sb$_2$O$_2$, WDS indicates that the average elemental ratio is Ba : Mn : Sb : O = 2: 2.85: 1.98: 1.89, with slight Mn and O deficiencies.

The crystal structures were identified using both power and single crystal x-ray diffraction (XRD) with Cu Kα (λ = 1.5418 Å) radiation, and neutron diffraction. For crystallographic and magnetic structure determination, single crystal neutron diffraction measurements were performed at the HB-3A four-circle diffractometer in the High Flux Isotope

Reactor (HFIR) at the Oak Ridge National Laboratory (ORNL). Neutrons of wavelength 1.003 Å were employed via a bent Si-331 monochromator without $\lambda/2$ contamination[16] to investigate $BaMn_2Sb_2$. Neutrons of wavelength 1.546 Å (with high flux involving ~1.4% $\lambda/2$ contamination) from the Si-220 monochromator[16] were used for $Ba_2Mn_3Sb_2O_2$. For each measurement, a crystal sample was placed on an Al holder with aluminum foil in an evacuated sample space. A high temperature close cycle refrigerator (CCR) furnace was installed, so that the measurements can be performed between 6 K and 800 K. Electrical resistivity and thermopower were measured using the standard four-probe method. To improve the electrical contact between the sample surface and leads, a thin layer of Au was deposited prior to attaching leads. Measurements of electrical resistivity, thermopower, and specific heat were performed in a Physical Property Measurement System (PPMS, *Quantum Design*) between 2 K and 400 K. Magnetic susceptibility measurements were carried out in a Superconducting Quantum Interference Device (SQUID, *Quantum Design*) between 2 K and 700 K, where a sample space oven option was used for high-temperature (> 400 K) measurements.

### III. RESULTS AND DISCUSSION

### A. Crystalline structure

The XRD measurements on crushed $BaMn_2Sb_2$ and $Ba_2Mn_3Sb_2O_2$ crystals show that there are no impurity phases, except for a tiny amount of residual Sn flux (not shown). The Rietveld analyses on the single crystal neutron diffraction data confirm the tetragonal structure with the space group *I4/mmm* (No. 139) for both compounds[17-18], as illustrated in Figs. 1(a) and 1(b), respectively. For both compounds, there is no evidence for any structural transitions down to 6 K. The refined lattice parameters, atomic positions, and reliability factors from our neutron data taken at 6 K are summarized in Table I. For $BaMn_2Sb_2$, the Mn(1)Sb$_4$ layer is similar to the FeAs layer formed with edge-shared tetrahedra. For $Ba_2Mn_3Sb_2O_2$, there are extra Mn(2)O$_2$ and Ba layers to separate Mn(1)Sb$_4$ tetrahedral layers, leading to a much larger lattice constant $c$ ( = 20.78 Å) than that of BaMn$_2$Sb$_2$ (= 14.28 Å). In contrast, the lattice constant $a$ for $Ba_2Mn_3Sb_2O_2$ is slightly shorter than that for BaMn$_2$Sb$_2$.

### B. Electrical resistivity and thermopower

The temperature ($T$) dependence of the in-plane resistivity ($\rho_{ab}$) of BaMn$_2$Sb$_2$ (circles) and Ba$_2$Mn$_3$Sb$_2$O$_2$ (squares) is shown in Fig. 2(a). For BaMn$_2$Sb$_2$, the resistivity increases with decreasing temperature in the entire temperature range measured, indicating a non-metallic behavior as reported previously [19]. Below $T_x \approx 180$ K, the change of $\rho_{ab}$ is slower than that at higher temperatures. We fit the electrical conductivity ($\sigma_{ab} = 1/\rho_{ab}$) above $T_x$ using the formula $\sigma_{ab} = \sigma_0 + B exp(-\frac{\Delta}{2k_B T})$, where $\sigma_0$ and B are constants, k$_B$ is the Boltzmann constant, and $\Delta$ is the energy gap. The fit yields $\sigma_0$ = 0.021 $\Omega^{-1}$cm$^{-1}$, B = 4830 $\Omega^{-1}$cm$^{-1}$, and $\Delta$ = 0.411 eV. We thus infer that BaMn$_2$Sb$_2$ a semiconductor. Below $T_x$, $\rho_{ab}(T)$ can be described by $\rho_{ab} = -6.47(5) \ln T + 77.2(1)$ $\Omega$-cm. This crossover from activated to logarithmic temperature dependence is very similar to that of Ba$_2$Mn$_2$Sb$_2$O[29]. The origin of the crossover in Ba$_2$Mn$_2$Sb$_2$O was attributed to an AFM phase transition. For BaMn$_2$Sb$_2$, it does not seem to have a magnetic phase transition around $T_x$, even though the magnetic susceptibility shows an anomaly.

In contrast, the electric resistivity of Ba$_2$Mn$_3$Sb$_2$O$_2$ differs from that of BaMn$_2$Sb$_2$, which (1) roughly three orders of magnitude lower, and (2) exhibits opposite temperature dependence. Note that, upon cooling, $\rho_{ab}$ initially decreases, followed by an anomalous increase below ≈320 K, and then decreases again below ≈260 K. The small upturn below $T_{N1}$ ≈320 K should be related to an AFM ordering in the Mn(1)Sb$_4$ layer (see discussion below). The non-monotonic character of $\rho_{ab}(T)$ suggests two-channel electrical conduction: one is based on the Mn(1)Sb$_4$ tetrahedral layer and another on the Mn(2)O$_2$ layer. The latter is more metallic than the former, leading to the metallic ground state (the residual resistivity is ≈35 mΩ cm). The metallic behavior observed in Ba$_2$Mn$_3$Sb$_2$O$_2$ is truly surprising, as all known 2322-type compounds are insulating[15]. The metallicity was only observed in Sr$_2$(Mn$_2$Cu)As$_2$O$_2$ [15], where Cu was considered to partially replace Mn in the MnAs$_4$ layer. Our results suggest that the metallic behavior in Ba$_2$Mn$_3$Sb$_2$O$_2$ is mainly driven by the presence of the MnO$_2$ layer. According to previous study, monolayer MnO$_2$ is non-metallic above ~60 K but metallic below ~60 K [20]. Such metal-insulator transition is absent in Ba$_2$Mn$_3$Sb$_2$O$_2$. There are two possible scenarios. One is that the Mn(2)O$_2$ layer in the Ba$_2$Mn$_3$Sb$_2$O$_2$ environment is intrinsically different from the free-standing monolayer, thus having different properties. This is further supported by the fact that the magnetic structure of the Mn(2)O$_2$ layer in Ba$_2$Mn$_3$Sb$_2$O$_2$ is different from that in bulk MnO$_2$ [21]. Another possibility is that the good conduction of the Mn(2)O$_2$ layer results from oxygen deficiency as reflected in the WDS measurements. However, WDS also indicates Mn deficiency. How Mn and O deficiency influences the electrical transport requires further investigation.

Despite the stark difference in the electrical transport between the two compounds, we observe very similar features in thermopower. Figure 2(b) displays the temperature dependence of thermopower ($S$) measured by applying thermal current along the *ab* plane for BaMn$_2$Sb$_2$ (circles) and Ba$_2$Mn$_3$Sb$_2$O$_2$ (squares), both having similar magnitude and undergoing several sign changes in the temperature range measured. The sign change clearly indicates that, in both compounds, there are two types of carriers: hole dominant when $S > 0$, and electron dominant when $S < 0$. For BaMn$_2$Sb$_2$, the minimum $S$ occurs near $T_x$. For Ba$_2$Mn$_3$Sb$_2$O$_2$, the sign change at $T_{N1} \approx 320$ K, and the peak $T_{N2} \approx 60$ K is likely associated with the magnetic transitions, as will be discussed below.

### C. Magnetic susceptibility and specific heat

The temperature dependence of magnetic susceptibility $\chi = M/H$ of BaMn$_2$Sb$_2$, measured at 1000 Oe applied magnetic field along both the *ab* plane and *c* axis, is shown in Fig. 2(c). Upon cooling, $\chi_{ab}$ initially increases with decreasing temperature, revealing a kink around 440 K as shown in the inset of Fig. 2(c), consistent with the previous report[19]. Below $T_x \approx 180$ K, $\chi_{ab}$ increases much faster than at high temperatures. Correspondingly, $\chi_c$ has a minimum at the same temperature, above which $\chi_c$ increases with temperature. The anomaly in the susceptibility was also reported previously but at a much lower temperature ~50 K[19].

To understand the nature of the anomaly occurring at $T_x$, we measured the temperature dependence of the heat capacity ($C_p$) for BaMn$_2$Sb$_2$, as shown in the inset of Fig. 2(b). There is no anomaly between 2 and 300 K, indicating the absence of a phase transition around $T_x$. At 300 K, the measured $C_p$ is 127 J/mol-K, which is consistent with the classical Dulong Petit specific heat value given by $C_p = 3nR \sim 125$ J/mol-K (where n is the number of atoms per formula and R is the molar gas constant). The obtained specific heat is quite similar to that of BaMn$_2$As$_2$ [22] and BaMn$_2$Bi$_2$ [23]. At low temperatures, $C_p(T)$ can be fitted by $C_p/T = \gamma + \beta T^2$, yielding $\gamma = 0.0064(1)$ J/K$^2$mol and $\beta = 0.0011(1)$ J/K$^4$mol. The small but nonzero Sommerfeld coefficient $\gamma$ suggests finite density of states likely resulting from impurity contribution to the electronic states. Assuming the $\beta T^2$ term arises from the lattice, we can estimate the Debye temperature $\theta_D$ using the relationship $\theta_D = (\frac{12\pi^4 Rn}{5\beta})^{1/3}$ [21], which gives $\theta_D = 207$ K. Note that BaMn$_2$Sb$_2$ is antiferromagntically ordered at low temperatures as shown below. Thus, the excitations of spin waves may also contribute to $\beta$ [24].

Interestingly, the temperature dependence of the magnetic susceptibility of $Ba_2Mn_3Sb_2O_2$ is quite different from that of $BaMn_2Sb_2$. As shown in Fig. 2(d), both $\chi_{ab}$ and $\chi_c$ of $Ba_2Mn_3Sb_2O_2$ increase with decreasing temperature, with a kink near $T_{N1} \approx 314$ K. Below $T_{N2} \approx 60$ K, both $\chi_{ab}$ and $\chi_c$ decrease with temperature before Curie tail-type increase. These anomalies suggest that there are two possible magnetic transitions.

### D. Magnetic structures determined by single-crystal neutron diffraction

To understand the complex physical properties observed in $BaMn_2Sb_2$ and $Ba_2Mn_3Sb_2O_2$, we carried out single-crystal neutron diffraction measurements. Figures 3(a) and 3(b) show the rocking curves of a nuclear (110) peak and a magnetic (101) peak of $BaMn_2Sb_2$ at different temperatures, respectively. The rocking curves are fitted by two Gaussian peaks that correspond to two single-crystal domains. Note that the intensity of the (110) peak, shown in Fig. 3(a), has weak temperature dependence between 6 and 680 K. In contrast, the (101) peak, displayed in Fig. 3(b), undergoes a rapid increase below 460 K. This indicates that there is a magnetic transition, and the (101) peak is not only a nuclear peak but a magnetic peak. We thus trace the temperature dependence of both the (110) and (101) peak intensities, which are plotted in Fig. 3(c). The sudden increase of the (101) peak intensity marks the magnetic transition at $T_N \approx 441$ K. It is worth noting that we did not observe the short-range magnetic order above $T_N$ since the linewidth and integrated intensity of the (101) peak are almost unchanged below/above $T_N$. Note that there is neither the intensity anomaly in a few peaks such as (110) (see dash line in Fig. 3(c)), (101), (002), (310) and (008) nor an emergence of new magnetic peaks below $T_x$, precluding a magnetic origin for the anomaly in the resistivity.

The refinement was performed by the Rietveld method using the FullProf package [25] on a set of nuclear and magnetic reflections at $T$ = 6 K for $BaMn_2Sb_2$. All the magnetic reflections can be indexed on the crystal unit cell with a magnetic propagation vector $k$ = (0, 0, 0). The SARAH representational analysis program [26] is used to derive the symmetry allowed magnetic structures. The decomposition of the magnetic representation into the irreducible representations is $\Gamma_3 + \Gamma_6 + \Gamma_9 + \Gamma_{10}$, two of which are FM ($\Gamma_3 + \Gamma_9$) and two are AFM ($\Gamma_6 + \Gamma_{10}$). The symmetry allowed basis vectors are summarized in Table II. The neutron diffraction pattern is fitted well using the $\Gamma_6$ model, indicating an AFM ordering below $T_N$. The determined magnetic structure is illustrated in Fig. 3(d), where Mn spins are antiparallel forming a nearest-neighbor (NN) antiferromagnetic alignment in both the *ab* plane as well as *c* axis, i.e., the *G*-type AFM order. The Mn moment is 3.83(3)$\mu_B$ pointing along the *c* axis, consistent with the sharper decrease of $\chi_c$ than $\chi_{ab}$ below $T_N$. The magnetic structure and the ordered moment are similar to those in $BaMn_2As_2$ [4] and $BaMn_2Bi_2$ [27]. It is worth noting that the $T_N$ of $BaMn_2Sb_2$ is significantly lower than that of $BaMn_2As_2$ (≈618 K) [4]. The reduced $T_N$ for $BaMn_2Sb_2$ may be attributed to the longer Mn-Sb distance of 2.753 Å as determined from our refinement, compared to the shorter Mn-As distance of 2.558(2) Å in $BaMn_2As_2$ [4], resulting in a weaker super-exchange (SE) magnetic interaction *via* the SE pathway Mn-Sb-Mn [10,28]. The ordered moment suggests $Mn^{2+}$ in its high spin state with total spin = 5/2, similar to that in $BaMn_2Bi_2$ [27] and $BaMn_2As_2$ [4]. The reduced moment for the high-spin state of $Mn^{2+}$ is common, likely resulting from the strong hybridization between Sb *p* and Mn *d* orbitals [10,29-30]. In addition, the slight Mn deficiency is also expected to reduce the actual moment.

To determine the magnetic structures of $Ba_2Mn_3Sb_2O_2$, two sets of data were collected at 100 K ($T_{N2} < T < T_{N1}$) and 6 K ($T < T_{N2}$), respectively. At 100 K, the magnetic peaks can be indexed in the unit cell, indicative of the propagation vector $\bm{k}$ = (0, 0, 0). Fig. 4(a) shows the (101) peak at indicated temperatures. Similar to what was observed in BaMn$_2$Sb$_2$, the intensity of (101) increases with decreasing temperature, while the (110) peak intensity remains almost unchanged (not shown here). This implies the same *G*-type AFM order as in BaMn$_2$Sb$_2$. The temperature dependence of the (101) peak shown in Fig. 4(d) identifies the clear magnetic transition at $T_{N1} \approx 314$ K, consistent with the kink observed in the resistivity (see Fig. 2(a)) and susceptibility (see Fig. 2(d)). Interestingly, new magnetic peaks appear with the propagation vector of $\bm{k}$ = (0.5, 0.5, 0) at 6 K. Fig. 4(b) shows the rocking curves for the magnetic peak (0.5 0.5 2) at 6 (black dots) and 100 K (red dots). The temperature dependence of the (0.5 0.5 0) magnetic peak shows another magnetic transition at $T_{N2} \approx 60$ K, corresponding to the anomaly in the susceptibility shown in Fig. 2(d). Symmetry-allowed magnetic structures with these two propagation vectors are summarized in Table II: all are AFM models.

Through the refinement, we obtain a *G*-type AFM order ($\Gamma_6$ model) for Mn(1) in the Mn(1)Sb$_4$ tetrahedral layer without moment on Mn(2) in the Mn(2)O$_2$ layer at 100 K, as illustrated in Fig. 4(e). The magnetic moment of Mn(1) also points along the *c* axis with $\bm{m_c}$ = 3.447(6)$\mu_B$ at 100 K. Such a *G*-type AFM order persists below $T_{N2}$, and the ordered moment slightly increases upon the decrease of temperature with $\bm{m_c}$ = 3.617(7)$\mu_B$ at 6 K. Similar to BaMn$_2$Sb$_2$, the *G*-type AFM order in the Mn(1) sublattice is formed by the SE magnetic interaction *via* the SE pathway Mn(1)-Sb-Mn(1). At 6 K, the refinement on 79 magnetic peaks with $\bm{k}$ = (0.5, 0.5, 0) reveals an AFM order in Mn(2) of the Mn(2)O$_2$ layer, with the refinement goodness $R_f$-factor $\approx$ 8.5 and $\chi^2 \approx$ 4.1. But the magnetic moment is along the diagonal [1 -1 0] direction, i.e., $\Gamma_3$ model. The NN Mn(2) spins are antiferromagnetically aligned in the *ab* plane but ferromagnetically aligned along the *c* axis, explaining why $\chi_c$ increases and $\chi_{ab}$ decreases below $T_{N2}$. The ordered moment is ≈ 3.56(4)$\mu_B$ for Mn(2), comparable to that of Mn(1), indicating both Mn(1) and Mn(2) are in the high spin state (S = 5/2). The magnetic structure of Ba$_2$Mn$_3$Sb$_2$O$_2$ at $T < T_{N2}$ is illustrated in Fig. 4(f), with the magnetic unit cell 4 times of the crystal unit cell. The AFM order in Mn(2) sublattice may be formed by the SE interaction via 180° Mn-O-Mn exchange pathway.

The lower $T_{N2}$ than $T_{N1}$ indicates that the dominant intralayer NN magnetic interaction in Mn(2) sublattice is weaker than that in the Mn(1) sublattice. In addition, the resolution-limited linewidths of the rocking curves of the magnetic (101) peak in Fig. 4(a) suggest a long-range *G*-type AFM order in the Mn(1) sublattice. The linewidths of the magnetic peaks with $k$ = (0.5, 0.5, 0) in the Mn(2) sublattice are much broader than the nuclear/magnetic (101) peak as shown in Fig. 4(b). While the H and K scans through the magnetic peaks are resolution limited, the L scan through magnetic peaks is much broader than the instrumental resolution, as illustrated in Fig. 4(c) for the (0.5 0.5 0) magnetic peak. This suggests a long-range order in the *ab* plane but a short-range order along the *c* direction. To determine the intrinsic correlation length along the L (or *c*) direction, we fit the experimental line-shape of the L scan to a Lorentz function $L(L) = \frac{c}{1+(L\times\xi)^2}$, convoluted with the instrumental resolution Gaussian function $G(L) = \exp(-L^2/(2\left(\frac{\sigma}{2\sqrt{ln4}}\right)^2))$. Here, $(\xi*2\pi/c)^{-1}$ is the half width at half maximum of the Lorentzian function where $\xi$ is defined as the magnetic correlation length[31,32], and σ is the full width at half maximum of the instrumental resolution Gaussian function with σ ≈ 0.0369 (rlu), i.e., 0.0112 Å$^{-1}$. The deconvolution yields the magnetic correlation length along the *c* axis $\xi_c$ ≈ 80 Å, confirming a short-range AFM order along the *c* axis for the Mn(2) sublattice. The perpendicular arrangement between the Mn(1) and Mn(2) moments as well as their distinct orders in Mn(1) (long-range order) and Mn(2) (short-range order) sublattices at $T < T_{N2}$ indicate that there is little coupling between them.

It is worthwhile comparing the magnetic structure of $Ba_2Mn_3Sb_2O_2$ with those of the isostructural $Sr_2Mn_3Sb_2O_2$ [11] and $Sr_2Mn_3As_2O_2$ [11,12]. In $Sr_2Mn_3Sb_2O_2$, a different AFM order ($\Gamma_5$) with the magnetic moment along the diagonal [110] direction was reported in the Mn(2) sublattice, although the Mn(1) sublattice exhibits the same G-type AFM order with the moment along the perpendicular c axis [11]. In addition, the Mn(2) sublattice in $Sr_2Mn_3Sb_2O_2$ has long-range AFM ordering. The difference may result from the shorter inter-Mn(2)$O_2$-layer distance in $Sr_2Mn_3Sb_2O_2$ (10.079 Å) relative to that in $Ba_2Mn_3Sb_2O_2$ (10.389 Å). For $Sr_2Mn_3As_2O_2$, a short-range AFM order in the Mn(2) sublattice was reported [11,12], with the same (G-type) AFM order in the Mn(1) sublattice. However, the magnetic structure of Mn(2) could not be determined because of the weak magnetic signal in a polycrystalline sample.

Let us now compare the magnetic structures of Mn(1)$Sb_4$ and Mn(2)$O_2$ layers with those of the Fe$As_4$ layers in Fe-based superconductors and $CuO_2$ layers in cuprate superconductors, respectively. The G-type AFM order with moment along the c axis in Mn(1)$Sb_4$ layers in both $BaMn_2Sb_2$ and $Ba_2Mn_3Sb_2O_2$ is distinct from the stripe-like AFM order with the ordered moment along the a axis in the orthorhombic structure, i.e., the diagonal [1 1 0] direction in the tetragonal notation in Fe-based superconductors. Within the $J_1 - J_2 - J_c$ Heisenberg interaction model [33], the in-plane checker-board-like AFM structure of the G-type order in $BaMn_2Sb_2$ and $Ba_2Mn_3Sb_2O_2$ suggests that the NN interaction $J_1$ is dominant, whereas the in-plane next-nearest-neighbor interaction $J_2$ is very weak or negligible. Thus, we conclude that $J_2 \ll J_1/2$ in $BaMn_2Sb_2$ and $Ba_2Mn_3Sb_2O_2$ (Mn(1)). In contrast, $J_2 > J_1/2$ leads to a stripe-like AFM order in Fe-based superconductors[34]. However, the in-plane magnetic structure in the Mn(2)$O_2$ layers of Ba$_2$Mn$_3$Sb$_2$O$_2$ is similar to that in the CuO$_2$ layers of La$_2$CuO$_{4+x}$[35] and La$_2$CuO$_{4-x}$[36]. The in-plane NN spins are antiparallel with an ordered moment along the diagonal [1 1 0] direction of the tetragonal notation, corresponding to the *a* (or *b*) axis in the orthorhombic notation in cuprates[35,36]. This comparison suggests that it is more desirable to manipulate the Mn(2)O$_2$ layer than the Mn(1)Sb$_4$ layer for superconductivity. It also demonstrates the significant effect of the intercalated Ba and Mn(2)O2 layers on the physical properties given the lattice symmetry is unchanged.

## IV. CONCLUSION

In summary, we have investigated the structural, magnetic, and electrical and thermal transport properties of BaMn$_2$Sb$_2$ and Ba$_2$Mn$_3$Sb$_2$O$_2$. BaMn$_2$Sb$_2$ is found to be a semiconductor with a *G*-type AFM order below $T_N \approx 443$ K. Interestingly, the addition of Ba and Mn(2)O$_2$ layers into BaMn$_2$Sb$_2$ result in a variety of rich physical properties in Ba$_2$Mn$_3$Sb$_2$O$_2$. Two magnetic transitions are observed in Ba$_2$Mn$_3$Sb$_2$O$_2$: (1) a *G*-type AFM order with moment along the *c* axis for Mn(1) in Mn(1)Sb$_4$ layers below $T_{N1} \approx 314$ K, and (2) an AFM order in the Mn(2)O$_2$ layers below $T_{N2} \approx 60$ K with moment along the diagonal direction in the *ab* plane. The perpendicular arrangement of two sets of magnetic moments suggests that their coupling is negligible. The metallic behavior observed in Ba$_2$Mn$_3$Sb$_2$O$_2$ may also be due to the Mn(2)O$_2$ layers, while the Mn(1)Sb$_4$ layers are non-metallic as seen in BaMn$_2$Sb$_2$. Remarkably, the Mn(2) sublattice forms a magnetic structure similar to that of cuprates, offering an opportunity for realizing superconductivity.

**Acknowledgments**: This work was primarily supported by the U.S. Department of Energy under EPSCoR Grant No. DE-SC0012432. A portion of this research used resources at the High Flux Isotope Reactor, a DOE Office of Science User Facility operated by the Oak Ridge National Laboratory.

**Table I.** Crystal structures of BaMn$_2$Sb$_2$ and Ba$_2$Mn$_3$Sb$_2$O$_2$ with the space group *I4/mmm* (no. 139) at 6 K determined by single crystal neutron scattering at HB-3A.

| Compound | atom | Wyckoff site | x | y | z | lattice constants | $R_f$-factor | $\chi^2$ |
|---|---|---|---|---|---|---|---|---|
| BaMn$_2$Sb$_2$ | Ba | 2a | 0 | 0 | 0 | a=b=4.39(1) | 4.9 | 1.25 |
| | Mn | 4d | 0 | 0.5 | 0.25 | c=14.28(5) | | |
| | Sb | 4e | 0 | 0.25 | 0.366(3) | | | |
| Ba$_2$Mn$_3$Sb$_2$O$_2$ | Ba | 4e | 0 | 0 | 0.417(2) | a=b=4.367 | 2.23 | 1.6 |
| | Mn(1) | 4d | 0 | 0.5 | 0.25 | c=20.779 | | |
| | Mn(2) | 2a | 0 | 0 | 0 | | | |
| | Sb | 4e | 0 | 0 | 0.168(3) | | | |
| | O | 4c | 0 | 0.5 | 0 | | | |

**Table II** The symmetry-allowed basis vectors [$m_x$ $m_y$ $m_z$] of the space group I4/*mmm* (no. 139) for the Mn(1) sublattice with **k** = (0, 0, 0) in BaMn$_2$Sb$_2$ and Ba$_2$Mn$_3$Sb$_2$O$_2$, and for Mn(2) sublattice with **k** = (0.5, 0.5, 0) in Ba$_2$Mn$_3$Sb$_2$O$_2$. Mn(1)$_1$: (0 0.5 0.25), Mn(1)$_2$: (0, 0.5, 0.75), and Mn(2):(0, 0, 0).

| **k** | atom | $\Gamma_3$ | $\Gamma_6$ | $\Gamma_9$ | $\Gamma_{10}$ |
|---|---|---|---|---|---|
| (0, 0, 0) | Mn(1)$_1$ | [0 0 $m_z$] | [0 0 $m_z$] | [$m_x$ -$m_y$ 0] | [-$m_x$ $m_y$ 0] |
| | Mn(1)$_2$ | [0 0 $m_z$] | [0 0 -$m_z$] | [$m_x$ -$m_y$ 0] | [$m_x$ -$m_y$ 0] |

| **k** | atom | $\Gamma_3$ | $\Gamma_5$ | $\Gamma_7$ |
|---|---|---|---|---|
| (0.5, 0.5, 0) | Mn(2) | [$m_x$ -$m_y$ 0] | [$m_x$ $m_y$ 0] | [0 0 $m_z$] |

Figures and captions:

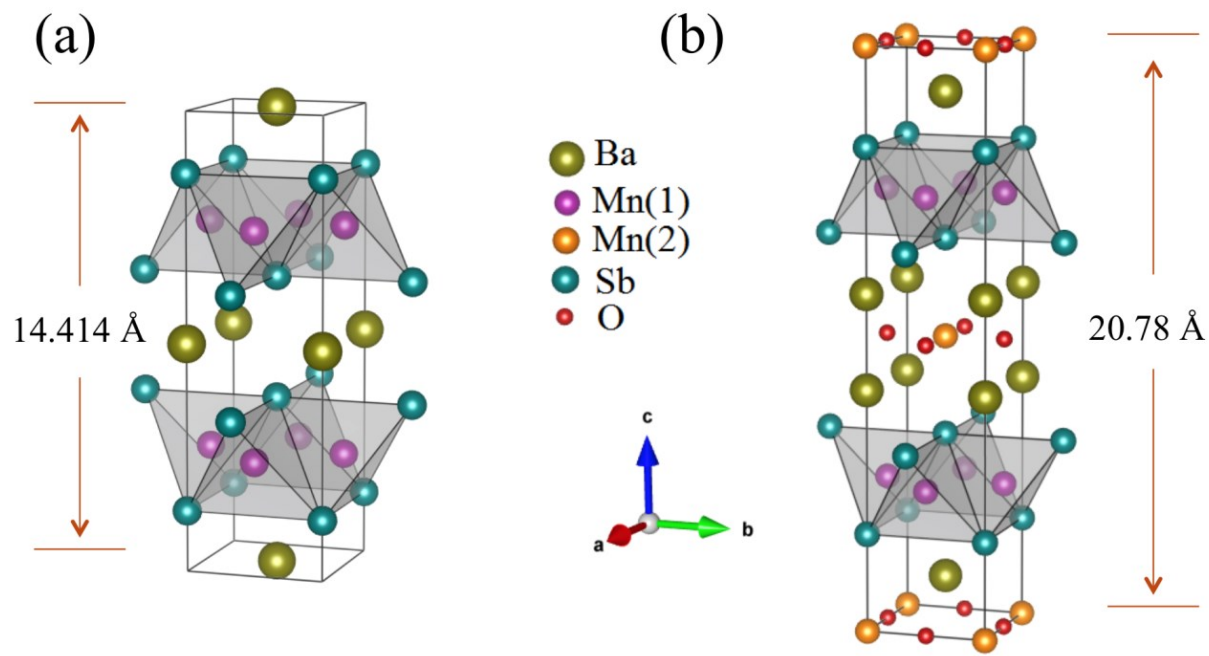

**Figure 1.** Crystal structure of BaMn$_2$Sb$_2$ (a) and Ba$_2$Mn$_3$Sb$_2$O$_2$ (b) using the best-fit parameters listed in Table I.

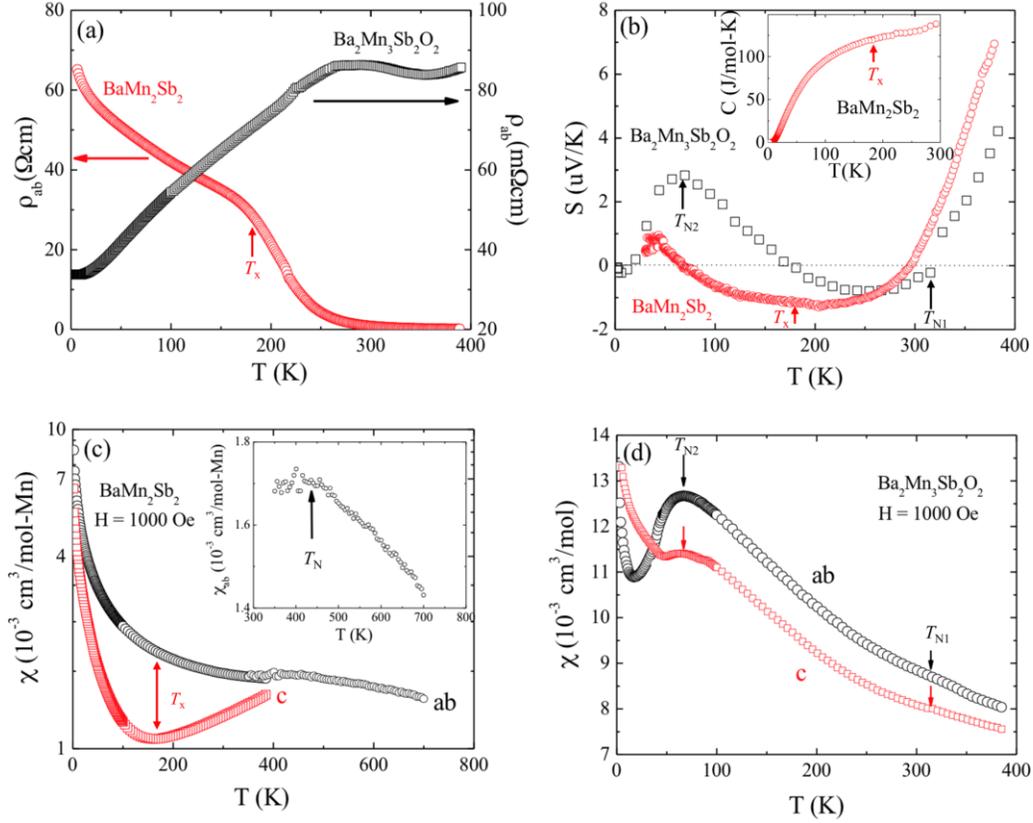

**Figure 2.** (a) Temperature dependence of the in-plane electrical resistivity, $\rho_{ab}$, for BaMn$_2$Sb$_2$ (circles) and Ba$_2$Mn$_3$Sb$_2$O$_2$ (squares); (b) Temperature dependence of Seebeck coefficient ($S$) of BaMn$_2$Sb$_2$ and Ba$_2$Mn$_3$Sb$_2$O$_2$. The inset shows the specific heat ($C_p$) as a function of temperature for BaMn$_2$Sb$_2$. (c) Temperature dependence of in-plane ($\chi_{ab}$) (circles) and out-of-plane ($\chi_c$) (squares) magnetic susceptibilities measured in a field of 1000 Oe for BaMn$_2$Sb$_2$. Inset: $\chi_{ab}$ between 300 and 700 K showing anomaly at $T_N \approx 441$ K; (d) Temperature dependence of $\chi_{ab}$ (circles) and $\chi_c$ (squares) for Ba$_2$Mn$_3$Sb$_2$O$_2$.

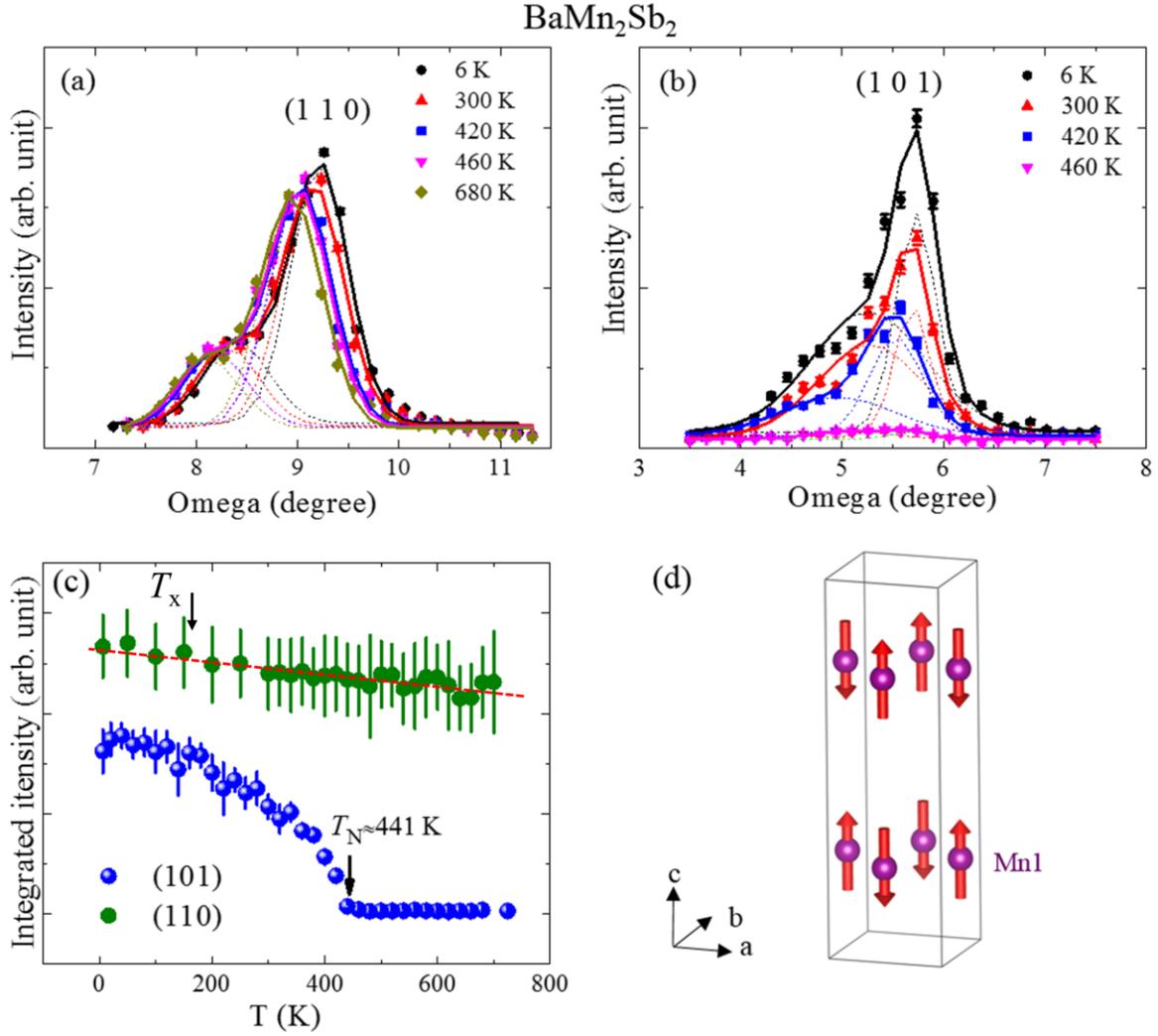

**Figure 3.** Neutron results on BaMn$_2$Sb$_2$: (a-b) Rocking curve scans at indicated temperatures for nuclear (110) peak (a) and magnetic (101) peak (b). The solid lines are fits using two Gaussian functions; (c) Temperature dependence of the integrated intensity of the (110) and (101) peaks obtained by the fits to the Gaussian function. The red dash line is the guide to eye; (d) Magnetic structure below $T_{N1}$.

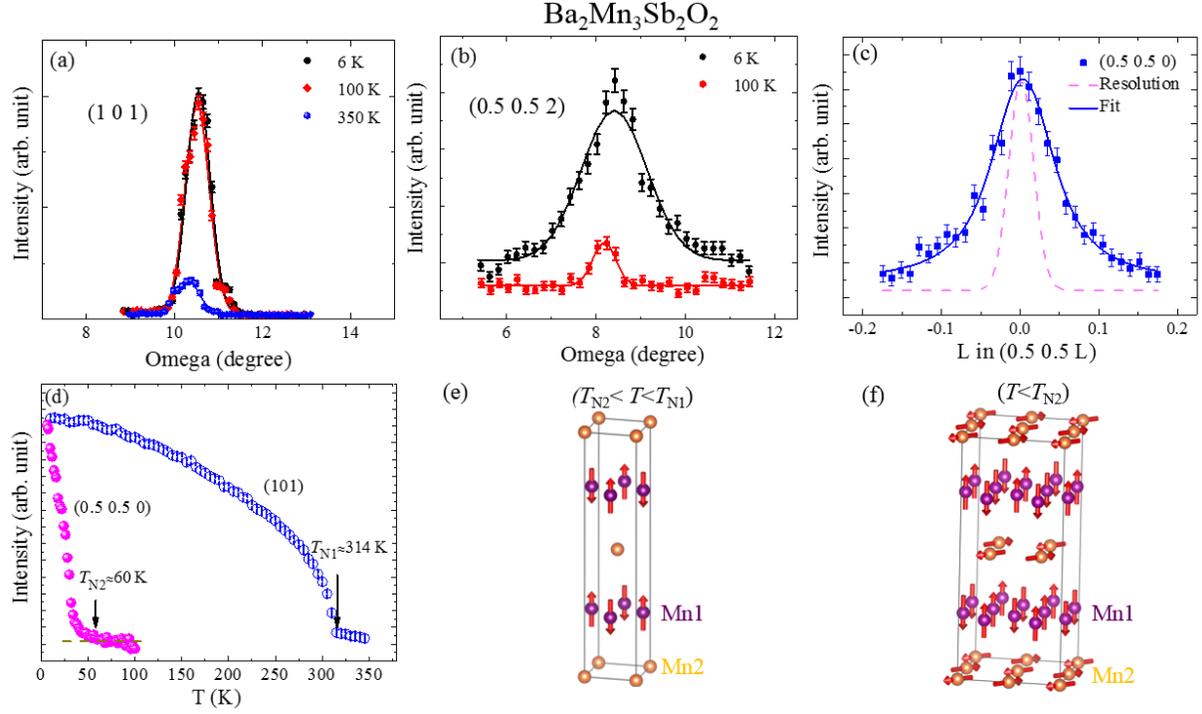

Fig. 4. Neutron results on $Ba_2Mn_3Sb_2O_2$: (a-b) Temperature dependence of the intensity for the nuclear/magnetic (101) peak (a), and the magnetic (0.5 0.5 2) peak (b). The solid lines are the fits to data using one Gaussian function. The weak peak at 100 K results from the λ/2 contamination. (c) L scan through the magnetic (0.5 0.5 0) peak at 6 K. The solid line is the fit to data using the Lorentz function convoluted with the instrumental resolution Gaussian function (dashed line). (d) Temperature dependence of the (101) and (0.5 0.5 0) peak intensities. $T_{N1}$ and $T_{N2}$ are indicated. (e-f) Magnetic structures in $T_{N2} < T < T_{N1}$ (e) and $T < T_{N2}$